\documentclass[11pt]{article}
\usepackage{authblk}
\usepackage{breakcites}
\usepackage{comment}
\usepackage{float}
\usepackage[margin=2.5cm]{geometry}
\usepackage{graphicx}
\usepackage{hyperref}
\usepackage[utf8]{inputenc}
\usepackage[version=4]{mhchem}
\usepackage{multirow}
\usepackage{upgreek}
\usepackage{url}
\usepackage[dvipsnames]{xcolor}

\title{Optimising experimental design in neutron reflectometry}
\date{}

\author[1]{James H. Durant}
\author[2]{Lucas Wilkins}
\author[1]{Joshaniel F. K. Cooper}
\affil[1]{\small{ISIS Neutron and Muon source, Rutherford Appleton Laboratory, Harwell Campus, OX11 0QX}}
\affil[2]{\small{School of Life Sciences, University of Sussex, Falmer, Brighton, BN1 9QG}}

\begin{document}
\maketitle

\begin{abstract}
Using the Fisher information (FI), the design of neutron reflectometry experiments can be optimised, leading to greater confidence in parameters of interest and better use of experimental time [Durant, Wilkins, Butler, \& Cooper (2021). \textit{J. Appl. Cryst.} \textbf{54}, 1100--1110]. In this work, the FI is utilised in optimising the design of a wide range of reflectometry experiments. Two lipid bilayer systems are investigated to determine the optimal choice of measurement angles and liquid contrasts, in addition to the ratio of the total counting time that should be spent measuring each condition. The reduction in parameter uncertainties with the addition of underlayers to these systems is then quantified, using the FI, and validated through the use of experiment simulation and Bayesian sampling methods. For a ``one-shot'' measurement of a degrading lipid monolayer, it is shown that the common practice of measuring null-reflecting water is indeed optimal, but that the optimal measurement angle is dependent on the deuteration state of the monolayer. Finally, the framework is used to demonstrate the feasibility of measuring magnetic signals as small as $0.01 \mu_{B}\slash \text{atom}$ in layers only 20\AA\ thick, given the appropriate experimental design, and that time to reach a given level of confidence in the small magnetic moment is quantifiable.
\end{abstract}

\section{Introduction}
Experimental design is a deep and open-ended subject, but it is one that can often be reduced to the quantitative problem of maximising the amount of information produced by a given experiment. It is often the case that the experimentalist has the ability to change a limited set of well-defined experimental conditions, and wishes to find those conditions that will reduce the uncertainty in the outcome by the largest amount. This is a particularly pertinent problem in the field of neutron reflectometry (NR), where the analysis of reflectivity data is ill-posed and, as a consequence, the quality of the data is of great importance. This is further compounded by the significant lead time and cost of accessing a neutron beamline; a typical experiment will take months to organise, last only a few dozen hours, and have operating costs of the beamline running into the thousands of pounds per day. Despite these significant challenges, NR can provide valuable insight. For example, investigating the structural properties of lipid leaflets in biological membranes \cite{Skoda2019}, probing magnetism in thin film heterostructures \cite{Liu2015}, examining surface chemistry at the air/water interface \cite{Welbourn2019}, and exploring layer structures in organic photovoltaics \cite{Zhang2017}.

Typically, the design of a NR experiment is determined by the knowledge gained in similar previous experiments, i.e., trial-and-error and ``rules of thumb''. To provide more rigorous and quantifiably optimised designs, we developed a framework \cite{Durant2021} using the Fisher information (FI) \cite{Fisher1925} that quantifies the maximum information in a NR experiment using the assumption of Poisson-based counting statistics. Using the FI and accurate experiment simulation, the framework was shown to be capable of optimising experimental design in a computationally-inexpensive manner. In this work, we extend the framework and demonstrate its utility in experimental design of a diverse selection of commonly-measured systems.

In NR, a collimated neutron beam is directed onto a surface of a sample and the intensity of the reflected radiation is measured as a function of angle and neutron wavelength, and for polarised measurements, spin state. Therefore, as an experimenter, one is presented with the non-trivial choice of angle(s) and counting time(s) for any NR measurement. The resulting reflectivity profile provides insight into the structure and properties of the sample surface including the thickness, scattering length density (SLD - the product of a material's density and its neutron scattering length), and interfacial roughness of any thin films. For certain systems, additional properties can be inferred, such as the sample hydration in liquid-based experiments and magnetic moments in magnetic samples; these systems often present added complexity in their experimental design, e.g., contrast choice.

In NR analysis, one is usually tasked with inversion of the reflectivity curve, with the aim of reconstructing the SLD profile. This is a known inverse problem due to the loss of phase information upon scattering \cite{Majkrzak1995} and consequently, the analysis is predominantly model-dependent. An initial model is typically defined using a series of contiguous layers, and is informed by prior knowledge of a system and the underlying science. The model reflectivity is calculated using the Abel{\`{e}}s matrix formalism for stratified media \cite{Abeles1948} or Parratt recursive method \cite{Parratt1954}. The difference between the model reflectivity and the data is then calculated, using a metric such as the chi-squared, and the model is then iterative modified in order to find the best agreement between the two. For this work, we use the Python packages \texttt{refnx} \cite{Nelson2019} and \texttt{Refl1D} \cite{Refl1D} for our model definitions and reflectivity calculations. Due to being an inverse problem, NR analysis is provably unsolvable in the general case with no additional information. Approximation methods have been devised to alleviate this issue and find likely solutions, including the use of evolutionary algorithms \cite{Storn1997} and even neural networks \cite{Mironov2021, Loaiza2021, Doucet2021, Greco2021}. However, there is no guarantee that such methods will provide the sample properties describing the ``true'' SLD profile. For a given reflectivity curve, there will be a large number of models which give equivalently good fits to the data. As a result, it is imperative that an experiment is well-designed to produce data that supports the experimentally-measured sample as best as possible.

For advanced data analysis, a Bayesian approach is commonly taken, often employing the use of sampling methods such as Metropolis-Hastings Markov Chain Monte Carlo (MCMC) \cite{Metropolis1953, Hastings1970} and nested sampling \cite{Skilling2004, Skilling2006}. These methods approximate parameter values and confidence bounds through estimation of the parameter posterior distributions. For this work, we use nested sampling implemented in the Python package \texttt{dynesty} \cite{Speagle2019}. In general, sampling methods will not be able to extract the maximum information that a data set contains, as calculated using the FI. They instead determine the maximum extractable information, given the data's limitations, e.g., parameter correlations which limit the ability to determine certain values. We use the computationally expensive sampling methods as a form of ``ground truth'' to validate the improvements in individual parameter variances and parameter-pair covariances using the experimental designs determined from the FI.

By simulating experiments with a known model, the FI framework can be used to calculate the information content of any set of experimental conditions without requiring expensive fitting, sampling calculations, or beamtime for data acquisition. The experiment simulation has been shown to be accurate, fast and general; any beamline and reflectometer can be simulated, given the instrument-specific incident flux profile. In our previous work, we used the framework to consider the experimental design of a common model system for structural biology: a 1,2-dimyristoyl-\textit{sn}-glycero-3-phosphocholine (DMPC) bilayer deposited onto a silicon surface. In this work, we apply a new optimisation approach, using the FI, to this system and several others that are frequently measured from different fields. We strive to answer some of the fundamental questions presented in the design of NR experiments including the choice of measurement angle(s), counting time(s), bulk water SLD(s) in liquid-submerged samples, and underlayer SLD(s) and thickness(es) in lipid bilayer and magnetic samples. These example model systems have been chosen to best demonstrate the wide applicability of the framework and can be modified to investigate other experimental conditions of interest. All examples shown here are presented on our GitHub repository \cite{GitHub}, alongside the underlying code for more advanced investigations.

\section{Methods}
\subsection{Reflectometry Models}
In NR, a model is defined by a structure representing a physical sample, some level of background noise, an experimental scale factor, and the instrument resolution function. The way this structure is defined is unimportant in our framework, as long as the model reflectivity can be calculated at a given neutron momentum transfer, Q.
$$Q=\frac{4\pi \sin{\theta}}{\lambda}$$
where $\lambda$ is the neutron wavelength and $\theta$ is the measurement angle. For this work, our model structures consist of contiguous layers of homogeneous SLD (i.e., slab models). For our experiment simulation, we use the Poisson-based counting statistics approach shown alongside the FI derivation in our previous work \cite{Durant2021}. As input, this method simply requires a model (or multiple models if multiple datasets are required), and the incident flux as a function of wavelength for the instrument being simulated. For this work, the flux profile (both polarised and non-polarised) was taken on the OFFSPEC neutron reflectometer \cite{Dalgliesh2011}. This flux profile is used with the model(s) to calculate the expected number of neutrons in each $Q$ bin, as is required for the FI. The framework also requires parameter values to be input which can be estimated from the given data or specified manually if known. We use \texttt{refnx} and \texttt{Refl1D} to define our models and to load the models' associated measured or simulated reflectivity data.

The simulation framework was developed on the principle that, for a given measurement angle, each neutron wavelength corresponds to a single momentum transfer point, $Q$, as is often the case in reflectometry. As a result, the experiment simulation would require modification for other techniques where this is not the case, e.g., small angle neutron scattering. Further, the simulation was developed to synthesise data without background subtraction. 

\subsection{Beamline Setup}
A conventional neutron reflectometer relies on two sets of apertures in order to collimate a neutron beam, limiting the beam footprint at the sample point. For the majority of the examples shown in this work it is appropriate to limit the beam footprint to be less than the sample extent, i.e. under-illuminating, so as not to add background noise by illuminating sealing gaskets or similar. When using an algorithm such as \cite{Cubitt2015}, the angular resolution of the measurement is dependent on the sample to detector distance and the detector pixel size. Therefore, there is an infinite set of aperture openings with equivalent resolutions that will achieve the same sample footprint. We chose the openings that maximise the flux on the sample, which can be proven to be half the maximum opening (which stays within the footprint) for each aperture. Thus, the beamlines here are already optimised (in hardware) for maximum flux, i.e., maximum information gain. For systems where this algorithm cannot be applied, the effect of resolution on the model (and therefore information) may need to be considered.

In this work, we consider the beamlines at the ISIS Neutron and Muon Source at the Rutherford Appleton Laboratory. These beamlines very commonly operate with a footprint of $\sim$60mm, as much of the sample environment equipment has been standardised. For the non-polarised experiments described in this work, we assume a footprint of 60mm (polarised experiments are described below). For all simulations, both apertures are scaled linearly with measurement angle to maintain a constant footprint on the sample. This scales the incident flux by a factor proportional to the angle squared and assumes no changes in beam profile with opening apertures; this is generally true in practice. Thus, when simulating an experiment, only a single incident flux profile is required, and the profile is scaled appropriately to the measurement time and angle.

It is not common practise on the ISIS beamlines to apply background subtraction, the preference normally being to fit the background in the analysis software. This is not the case in all reflectometer beamlines, and many rely on background subtraction for all of their measurements so we mention some considerations which should be taken into account when using this framework. The issue lies in the fact that the statistics governing background subtracted data are no longer Poisson, but follow something like a Skellam distribution (i.e., the difference between two Poisson distributions). In this case our derivation of the FI no longer holds since equations 5 and 6 in \cite{Durant2021} are not correct for anything other than a Poisson distribution. The best way to circumvent this issue is to include the background in the model as a parameter; our framework then still applies, and indeed from a statistics point of view, it is a more correct procedure. From a practical standpoint, we believe that given the uncertainties associated with model parameters before they are measured (i.e., in the experimental design phase) a sufficiently optimal experiment may still be designed solely using a modification to the incident flux profile and its behaviour at higher $Q$.

\subsection{Experimental Design Optimisation} \label{optimisation}
The derivation and initial demonstration of the FI for neutron reflectometry is given fully in \cite{Durant2021} and will not repeated here. Instead, we solely describe how the systems being investigated are described, and how the FI is utilised in optimising their experimental design.

By refining the ideas introduced in our initial work, we have developed an improved methodology for experiment optimisation using the FI. Previously, we had not been able to directly compare the information content of parameters of differing units, since parameters can be on widely differing scales and the units of the FI are \textit{nats} (natural unit of information) per parameter unit squared. We addressed this issue by weighting parameter importance. We have assumed that our model parameters are all of equal importance in terms of their specified units, but the methodology can be easily adapted to accommodate a custom importance weighting if desired. Another limitation of our previous work was that we had only considered the diagonal entries of the FI matrix (the information content of each individual parameter). With all parameters now on the same scale, we can include the off-diagonal elements of the FI matrix by finding the experimental conditions that maximise the minimum eigenvalue: a maximin approach. This effectively finds the conditions that improve the ``worst'' possible combination of parameters the most, thereby minimising the overall uncertainty of experiment. The reader is referred to the supporting information (SI) for further details of this approach and the importance scaling.

One benefit of the new optimisation method is that the optimisation function is multivariate scalar, whereas our previous approach required optimisation of a vector of parameter information values. Bounds are placed on the variable experimental conditions and so the optimisation is constrained. It is therefore possible to take every experimental condition into account simultaneously and find the ``complete'' optimal experimental setup using established optimisation algorithms, e.g., differential evolution (DE) \cite{Storn1997}; for a minimising algorithm like DE, the negative of the minimum eigenvalue can be used as the optimisation objective. This ``complete'' optimisation can account for any complex relationships between experimental conditions that may be missed if these conditions were considered individually.

\subsection{Lipid Bilayers} \label{bilayer_methods}
In this work, we take two lipid bilayer models and investigate the choice of measurement angles, counting times, contrasts and the effect of underlayers on the models' parameter uncertainties. The first of these models is taken from our previous work where we quantified the information content of each model parameter of a DMPC bilayer deposited onto a silicon surface as a function of the bulk water SLD. The bilayer model was defined by two lipid leaflets with fixed surface coverage. The lipids were measured against two water contrasts, \ce{H2O} and \ce{D2O}, using the CRISP neutron reflectometer \cite{Penfold1987} as part of the ISIS neutron training course. The data were simultaneously fitted using \texttt{RasCAL} \cite{RasCAL} with the fitting constrained against measured data for a bare \ce{Si}/\ce{D2O} interface including a native \ce{SiO2} layer. The fitted model was converted to \texttt{refnx} and reparameterised as a function of the bulk water contrast SLD, enabling simulation of the bilayer on the OFFSPEC reflectometer with arbitrary contrast SLD. Further details of the model parameterisation can be found in our previous work \cite{Durant2021}.

In addition to the DMPC bilayer, we also consider a more sophisticated bilayer model in this work. We take a data set from \cite{Clifton2016} that studied a highly asymmetric bilayer structure made of a phospholipid-rich inner leaflet composed of 1,2-dipalmitoyl-\textit{sn}-glycero-3-phosphocholine (DPPC), and a Ra lipopolysaccharide (LPS) outer leaflet. Specifically, we investigate tail-deuterated DPPC (d-DPPC) data. Three isotopic contrasts were measured for the DPPC/RaLPS bilayer using the INTER reflectometer \cite{Webster2006}: d-DPPC/RaLPS in \ce{D2O}, d-DPPC/RaLPS in silicon-matched water (SMW), and d-DPPC/RaLPS in \ce{H2O}. As with the DMPC bilayer, the model for this data was originally defined and fitted with \texttt{RasCAL} but was recreated in \texttt{refnx} for this work. The model accounted for each individual layer hydration, the presence of bilayer defects across the surface and the lipid asymmetry. For additional details of the model parameterisation, the reader is directed towards the SI where the fitted model parameters for both models can also be found.

Like our previous work, we have assumed that the molecular volumes do not vary with measurement conditions. These volumes may not necessarily be constant in practice \cite{Campbell2018}, but to simplify the models, the volumes have remained fixed. Using this assumption, we can optimise the choice of contrast(s) for both bilayer models. To support the addition of underlayers, whose presence changes the reflectivity curves and may improve the information gain, we set the \ce{SiO2} hydration to 0\%. We then add a layer of given SLD and thickness between the \ce{SiO2} and the inner bilayer headgroups for each underlayer to add; a fixed 2\AA\ roughness and 0\% hydration is assumed for each underlayer. With the addition of these underlayers, the \ce{SiO2} hydration and \ce{SiO2}/bilayer roughness parameters are excluded from the FI calculation. 

When simulating data for both bilayer models, angles of $0.7^{\circ}$ and $2.3^{\circ}$, and times of 15 and 60 minutes, respectively, are used (or the same ratio of times between angles is used when the total time is altered) with 100 data points per angle. The experimental scale factor and instrument resolution function used is 1.0, $5 \times 10^{-6}$, and constant 2\% $dQ/Q$ respectively. The background level varied between $2\times 10^{-6}$, for \ce{D2O}, and $4\times 10^{-6}$, for \ce{H2O}. We optimise the measurement angles over the interval $[0.2, 4.0]^{\circ}$, representing realistic bounds on the physically possible measurement angles of the OFFSPEC reflectometer being simulated. The contrast SLDs, underlayer thicknesses and underlayer SLDs are optimised over the intervals $[-0.56, 6.36] \times 10^{-6} \text{\AA}^{-2}$ (pure \ce{H2O} to pure \ce{D2O}), $[0, 500]$\AA\ and $[1, 9]\times 10^{-6} \text{\AA}^{-2}$ respectively, the latter being the range of SLD's which most non-isotropically enriched film materials would fall.

We first consider a simple form of optimisation where we visualise the optimisation space for the simultaneous choice of two contrasts for both bilayers, assuming no prior measurements and assuming that the two contrasts are measured for equal amounts of time. We then validate the improvement attained with the optimal solution, suggested by the framework, using nested sampling. Following this, we generalise by optimising the choice of up to 4 contrasts and additionally optimise the proportion of the total time spent measuring each contrast. However, as we go beyond optimising two experimental conditions at once (e.g., three or more contrasts), we run into two problems: a brute-force approach becomes computationally difficult or infeasible, and the optimisation space becomes challenging to visualise. Therefore, for our more advanced optimisation results, we apply the DE algorithm. To significantly reduce the dimensionality and complexity of the optimisation problem, we make the simplifying assumption that all contrasts are measured using the same angles and the same proportion of times between angles (but the total time spent measuring each contrast is not necessarily equal). In practice, this may not always be the case. However, for these model systems, it is unlikely the added complexity of considering different angles (and counting time splits between angles) for each contrast would influence the end results drastically. Using the simplifying assumption, we optimise contrasts and then separately optimise the measurement angles (and proportion of the time spent measuring each angle) for the contrasts.

Thus far, we have considered the choice of measurement angles, counting time ratios and contrasts for the bilayer models but we also have the ability to improve the experiment through modification of the sample structure itself. When a chemical surface other than silicon dioxide is desired to interact lipids with it is common practise to use a gold layer grown onto the silicon. This can be functionalised almost arbitrarily using self-assembled thiol-based monolayers to give any desired chemical termination. In addition to chemical functionalisation there is also a portion of the field using magnetic underlayers as an additional contrast mechanism. Commonly, a permalloy or nickel thin film is deposited onto the substrate, and then coated in gold again with a functionalised surface. The whole system is then measured with a polarised neutron beam, using the different SLD's the two spin states experience in the permalloy either instead of changing the water contrast, or in addition to it. Such systems offer a wealth of potential optimisation, however, the complexity of them means that the benefits of the magnetic contrasts will be highly model and even beamline specific. Thus, we do not investigate them further in this work, but do expect the tools described here to have huge utility in designing such experiments. We instead demonstrate the ability to increase experimental information solely through the addition of an underlayer or underlayers whose SLDs and thicknesses are optimised using DE. We also visualise the optimisation space for a single added underlayer, compare the results to gold and Permalloy underlayers commonly used \cite{Clifton2015}, and use nested sampling to validate the improvement in the estimated parameter uncertainties and posterior distributions. We note that when investigating the parameter uncertainties, we do not include the underlayer parameters, since they are essentially "nuisance" parameters, whose values are needed, but not important for the comparison. In real experiments the underlayer parameters are likely to be well defined enough by the initial characterisation of the wafer, (due to strong scattering and a large number of fringes) that very little additional ambiguity will be included in the bilayer parameters.

When competing experimental conditions are involved, we perform our optimisation using DE and assuming a fixed time-budget; when optimising contrasts (or angles), the time spent measuring all contrasts (or angles) is constrained to be constant and the proportion of the total time spent measuring each contrast (or angle) is optimised in addition to the contrast SLDs (or angles) themselves. The optimisation is also performed using a large total time-budget is used to reduce the impact of noise in the simulated data and assist in convergence to the global optimum. To illustrate this, suppose we were optimising two contrasts and DE was in a region of the optimisation space where one contrast had a counting time of 1\% and the other 99\% of the total time budget, and the total time budget was only 10 minutes, the former contrast would only have a counting time of 6 seconds resulting in very poor data and thus hindering the optimisation process. It is worth noting that using a large total time does not influence the results of optimisation (other than reducing the effect of noise); if the total time is increased and the ratio of times between contrasts remains fixed, there is no change in the relative difference in information between contrasts.

\subsection{Kinetics} \label{kinetics_methods}
Often in NR, we wish to make measurements where the sample changes over time, for example, due to a sample degrading due to some reaction. Typically, when measuring these kinetic systems, only a single measurement is possible, and the designs of such experiments must reflect this. Therefore, when considering such a design, a time budget becomes irrelevant as there are no longer multiple conditions to consider splitting a time budget between. Instead, we wish to make the measurement that results in the greatest information about the system using the experimental conditions we do have control over; for a liquid-submerged sample, these are the measurement angle and liquid contrast.

To demonstrate the framework's flexibility in the experimental design of a kinetic system, we modelled a system from \cite{Clifton2011} that investigated the binding of puroindoline-a (Pin-a) proteins to lipid monolayers composed of 1,2-dipalmitoyl-sn-\textit{glycero}-3-phospho-(1-rac-glycerol) (DPPG); the original measurements were carried out using the SURF neutron reflectometer \cite{Penfold1997} with the isotopic contrasts measured being equilibrium Pin-a adsorbed d-DPPG (chain-deuterated DPPG) monolayer on null-reflecting water (NRW), h-DPPG (hydrogenated DPPG) on NRW and h-DPPG on D2O. The model was originally defined and fitted \texttt{RasCAL} but for this work, we have recreated the model in \texttt{refnx} without a protein, leaving just a monolayer. In the original work, the model SLDs and thicknesses were fitted directly and then converted to volume fractions. We use a more sophisticated model here defined by area per molecule (APM). The model can describe both the h-DPPG and d-DPPG, accounts for the headgroups containing water through defects across their surfaces and also the water bound to the hydrophilic headgroups. The model has been parameterised as a function of the bulk water contrast SLD to facilitate contrast optimisation. The full details of the model parameterisation and fitting can be found in the SI. As with section \ref{bilayer_methods}, we have assumed the molecular volumes are invariant with changing experimental conditions.

To introduce kinetics into the model, we simulate the surface excess decreasing over time by increasing the lipid APM, since the surface excess (in $\text{mg} \ \text{m}^{-2}$), $\Gamma$, is related to the APM (in $\text{\AA}^{2}$) by
$$\Gamma = \frac{\text{Molecular Weight} \times 10^{23}}{N_{A} \times \text{APM}}$$
Typically, the reflectivity profile of a monolayer is fairly featureless and so the model parameters are not particularly well-defined when only considering a single contrast (as is the case for our model system). As consequence, we only investigate the parameter of greatest significance: the lipid APM. As there is just a single parameter, there are no parameter-pair covariances to consider and so we can simply use the FI itself as our objective to maximise.

For our results, we simulate data using an experimental scale factor of $1.0$, instrument background of $5 \times 10^{-6}$ and resolution function of constant 2\% $dQ/Q$. To account for the sample degradation, data is simulated for 20 lipid APM values, ranging from the fitted value of $54.1 \text{\AA}^{2}$ to $500 \text{\AA}^{2}$, with the FI calculated over the entire simulated data set; this quantifies the maximum information obtainable about the APM over the full experiment using the given conditions. The entire data set is given a time-budget of 150 minutes with 100 data points per APM value. When optimising the experimental design, the measurement angle and contrast SLD are optimised over the intervals $[0.2, 4.0]^{\circ}$ and $[-0.56, 6.36] \times 10^{-6} \text{\AA}^{-2}$ respectively.

\subsection{Magnetism} \label{magnetism_methods}
An area where it is particularly important to be able to discern small signals and maximise differences between models is in thin film magnetism. It is often the case that small moments are induced by a ferromagnet in neighbouring layers, proximity magnetism \cite{Khaydukov2013, Cooper2017, Duffy2019, Zhan2019, Inyang2019}, or where a layer might only have a very small moment in the first place, but the magnitude of this moment is particularly important to know. The advantage of neutrons in these situations is that they provide an absolute moment, so in many ways are the idea tool for the task. However, neutron sensitivity is much lower than some other techniques, with a rough rule of thumb putting the minimum measurable moment with polarised neutron reflectivity at around 0.05 $\mu_{B}\slash \text{atom}$. 

In this work, we use a data set from \cite{Cooper2017} that was measured to quantify the moment that a yttrium iron garnet (YIG) film, grown on a yttrium aluminium garnet (YAG) substrate, induces in an adjacent platinum capping layer. This experiment was already unusually sensitive to the induced moment, allowing a limit of $\pm 0.02 \mu_{B} \slash \text{atom}$ to be placed on the induced magnetism. The model for this work was originally defined and fitted in the reflectivity analysis package \texttt{GenX} \cite{Bjorck2007} but was recreated in \texttt{Refl1D} for our experimental design analysis. The polarised, but not analysed, measurements were made using the MAGREF neutron reflectometer \cite{Ambaye2008}. The reader is directed towards the SI for further details of the model parameterisation and fitting.

To keep the focus of the experimental design on the sample itself, we use three pre-defined incident angles of $0.5^{\circ}$, $1.0^{\circ}$, and $2.0^{\circ}$ with measurement times of 30, 60 and 120 minutes respectively (or the same ratio of times between angles when the total time is altered), and 100 data points per angle: a commonly used array of angles and times for OFFSPEC when polarised. More angles are required than for unpolarised measurements since polarising systems reduce the incident flux at low wavelengths. As such, a different incident flux profile for OFFSPEC is used for this section, suitable for simulating polarised data. For our simulation, we use an experimental scale factor, level of background noise and instrument resolution function of 1.0, $5 \times 10^{-7}$, and constant 2\% $dQ/Q$ respectively. We assume a scale factor of unity, i.e., every incident neutron interacts with the sample, but this is usually not the case for magnetic samples due to the difficulties of getting large homogeneous films. For scale factors less than unity, the results shown here will hold true, but the counting times will scale inversely to the scale factor (e.g., a scale factor of 0.1 would require counting 10 times longer to achieve the same uncertainty).

Without changing any of the interface physics, we are able to reparameterise the model as a function of the thicknesses of both of the YIG and platinum layers in a quest to make ourselves more sensitive to the moment. For simplicity we assume that there is a constant moment of $0.01 \mu_{B} \slash \text{atom}$ (equivalent to a magnetic SLD of $0.0164 \times 10^{-6} \text{\AA}^{-2}$) in the 21\AA\ thick platinum layer, i.e., within our current error bound. When changing the platinum layer thickness, we constrain the magnetic moment to remain within this 21\AA\ layer as not to increase the total magnetism in the system as the platinum layer thickness increases. We calculate the optimal thicknesses of the YIG and platinum layers in the intervals [400,900]\AA\ and [20,100]\AA\ respectively.

To validate the improvement with the suggested experimental design, we use the ratio of likelihoods between two models, one with an induced moment of 0.01 $\mu_{B}\slash \text{atom}$ in the platinum layer and one with no moment, as a function of measurement time, to determine what level of statistics are required for a differentiable difference between the two models. The likelihood, $\mathcal{L}$, provides a measure of difference between a given data set and model and, for this work, is defined as
$$
\ln{\mathcal{L}} = -\frac{1}{2} \sum_{i=1}^{N} \Bigg(
                                                    \bigg( \frac{r_i - r_{i_m}}{\delta r_i} \bigg)^2 + 
                                                    \ln{\big[ 2\pi(\delta r_i)^2 \big]} 
                                               \Bigg)
$$
where, $N$ is the number of data points, $r_i$ is the experimental/simulated reflectivity at the $i$th $Q$ point, $\delta r_i$ is the uncertainty in the experimental/simulated reflectivity at the $i$th $Q$ point and $r_{i_m}$ is the model reflectivity at the $i$th $Q$ point calculated using the Abel{\`{e}}s matrix formalism. The data for the models is simulated with the induced moment and hence we should expect the correct model to become more ``likely'' as the measurement time is increased. This process is performed twice, once with an optimised design and once with a sub-optimal design, to illustrate how the optimised design reduces the time to confidently discern the magnetic moment.

\section{Results and Discussion}
\subsection{Lipid Bilayers}
\subsubsection{Contrasts} \label{bilayer_results_1}
\begin{figure}
\centering
\includegraphics[width=0.8\textwidth]{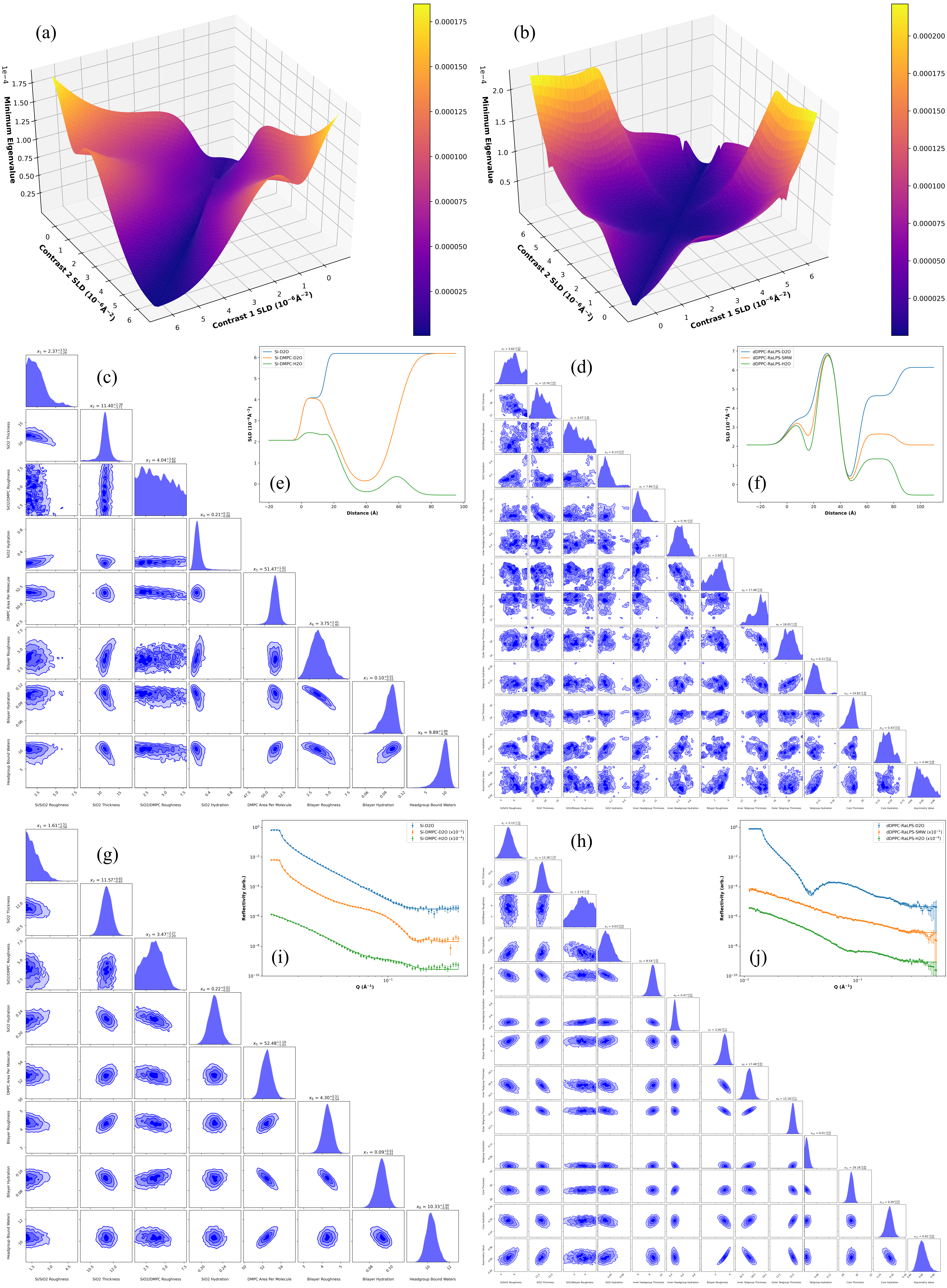}
\caption{Plots on the left correspond to the DMPC bilayer and plots on the right to the DPPC/RaLPS bilayer. Shown (\texttt{a} and \texttt{b}) are the plots of minimum eigenvalue versus bulk water contrast SLD for the bilayer models, for the simultaneous choice of two contrasts assuming no prior measurement. Higher values of minimum eigenvalue are better. Also shown are the nested sampling corner plots from sampling simulated data of solely \ce{D2O} (\texttt{c} and \texttt{d}) and simulated data of \ce{D2O} and \ce{H2O} (\texttt{g} and \texttt{h}). Insets are the fitted SLD profiles (\texttt{e} and \texttt{f}) and reflectivity curves (\texttt{i} and \texttt{j}) for the experimentally-measured data sets of the two models. For clarity, the \ce{H2O} and \ce{D2O} DMPC bilayer data sets have been offset by factors of $10^{-2}$ and $10^{-4}$ respectively. The \ce{H2O} and SMW DPPC/RaLPS data sets have been offset by the same factors, respectively. All figures are generated by scripts available the GitHub \cite{GitHub}, with full resolution figures stored there also.}
\label{fig:bilayer_contrasts}
\end{figure}

The fitted SLD profiles and experimental reflectivity data for the two lipid bilayer systems of section \ref{bilayer_methods} are shown in figure \ref{fig:bilayer_contrasts}. Figure \ref{fig:bilayer_contrasts} also visualises the optimisation space for the simultaneous choice of two contrasts, for the two models, assuming no prior measurement. The plot has symmetry as, assuming that each contrast is measured under the same conditions, the choice of contrast is commutative (the order does not matter). For both bilayer models, it is clear that measuring contrasts of maximum difference in SLD is optimal and that measuring the same contrast twice (i.e., for twice as long), or contrast matching a layer are both sub-optimal. These results agree with both common practice and sampling methods. This is illustrated in figure \ref{fig:bilayer_contrasts} where the nested sampling corner plots (using the default \texttt{dynesty} stopping criteria) from sampling simulated data of \ce{D2O} and \ce{H2O}, and \ce{D2O} measured for twice as long are shown. The distributions are clearly much better defined (i.e., more Gaussian) in the \ce{D2O} and \ce{H2O} case when compared to only \ce{D2O}. In addition, 95\%/2-sigma credible interval of the parameters are considerably lower in the former case when compared to the latter, with the median parameter error at $\approx50\%$ of the parameter value for only \ce{D2O} decreasing to $\approx10\%$ with the addition of the second contrast.

\subsubsection{Time-constrained Optimisation} \label{bilayer_results_2}
Table \ref{tab:bilayer_contrasts} shows the optimised contrasts and counting time splits for each bilayer model using 1 to 4 contrasts, assuming a fixed time-budget. For a single contrast, it appears that \ce{D2O} is optimal for DMPC and \ce{H2O} optimal for DPPC/RaLPS. Since the DMPC bilayer tailgroup is hydrogenated and the DPPC tailgroup deuterated, we hypothesise that the contrasts are optimal as a result of maximum contrast variation. For both models, there is a clear and large improvement in measuring two contrasts over just one. This is particularly noticeable for the DPPC/RaLPS model as it is defined using a large number of model parameter which are poorly described with just a single contrast. When measuring two contrasts, the results clearly agree with those of figure \ref{fig:bilayer_contrasts}: measuring \ce{D2O} and \ce{H2O} is optimal for both models, given the measurement conditions detailed in section \ref{bilayer_methods}.

\begin{figure}[t]
\centering
\includegraphics[width=0.95\textwidth]{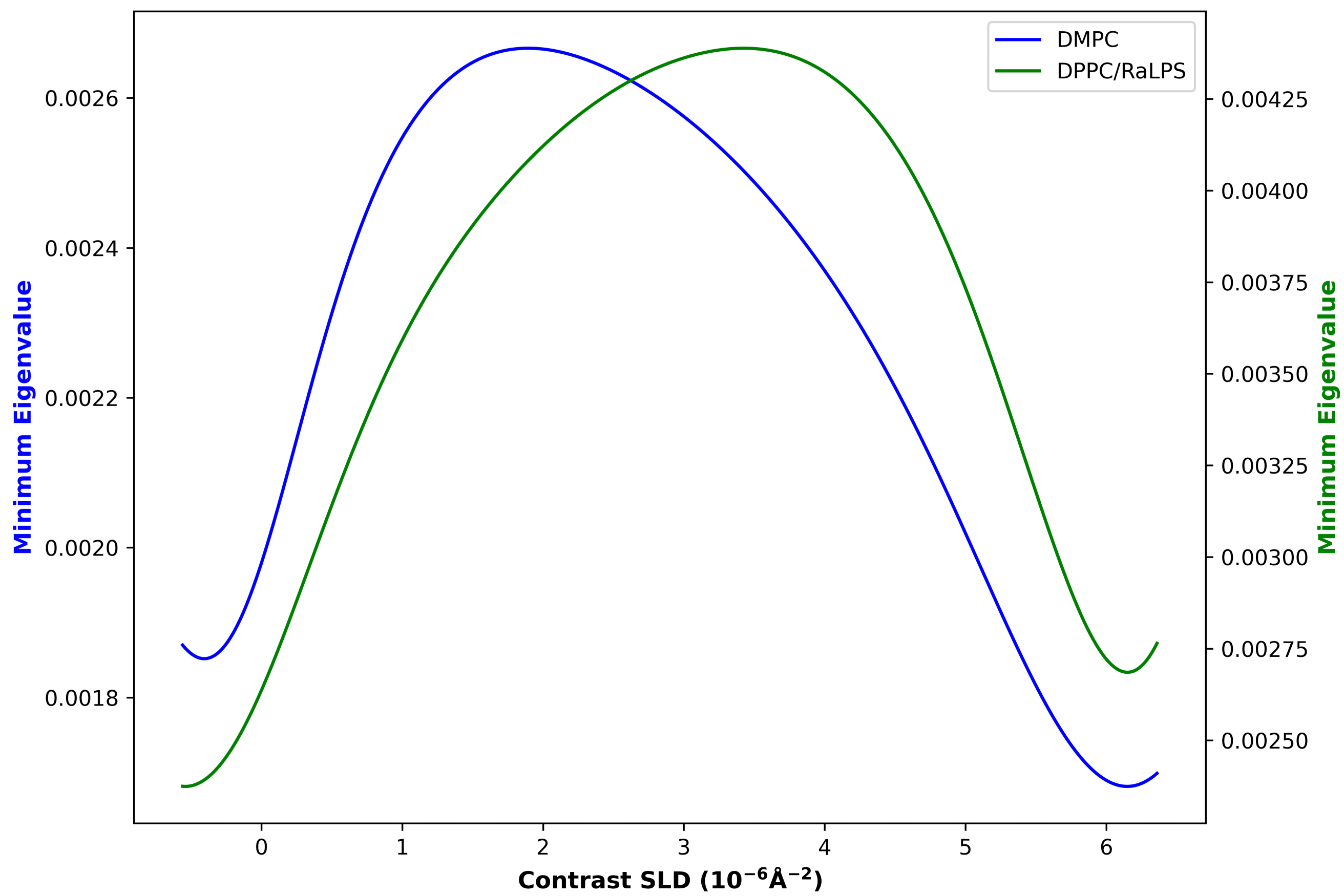}
\caption{Minimum eigenvalue versus bulk water contrast SLD for the DMPC (blue) and DPPC/RaLPS (green) bilayer models, for a third contrast choice, assuming \ce{D2O} and \ce{H2O} have been previously measured.}
\label{fig:bilayer_third_contrasts}
\end{figure}

When considering three contrasts, it can be seen that a small improvement can be obtained with an additional contrast of SLD around $1.9 \times 10^{-6} \text{\AA}^{-2}$ for DMPC and $3.4 \times 10^{-6} \text{\AA}^{-2}$ for DPPC/RaLPS. This conforms with current practices where \ce{D2O} and \ce{H2O} are typically measured and a third contrast of SMW (SLD of $2.07 \times 10^{-6} \text{\AA}^{-2}$) is sometimes also measured. Although these contrast SLDs differ from SMW, an improvement can still be attained from measuring SMW. This is illustrated in figure \ref{fig:bilayer_third_contrasts} which shows how the minimum eigenvalue changes with third contrast SLD for both models; the split of the total counting time between each contrast was defined by the three contrast results of table \ref{tab:bilayer_contrasts}. We emphasise that the minimum eigenvalues should not compared between models (e.g. DMPC vs DPPC/RaLPS), since they have different number of parameters, but rather the water contrast at which they are individually maximised is of importance. When considering which third contrast improves both models the most, SMW is an ideal candidate and likely a suitable choice if the optimal (model-dependent) third contrast is unknown. It is worth noting that gains achieved in measuring a third contrast are relatively small and the time to change contrast may outweigh the minor gains achieved when compared to measuring either \ce{D2O} or \ce{H2O} for longer. Therefore, we proceed with the assumption that \ce{D2O} and \ce{H2O} will be measured. Finally, the results suggest that measuring more than three contrasts is unnecessary given four or more contrasts, the solution is essentially the same as with three after combining the times of (near) identical contrast SLDs.

From table \ref{tab:bilayer_contrasts} it can also be seen that the time spent measuring each contrast is highly model-dependent; for the DMPC model, a much smaller proportion of the total time budget need be spent measuring \ce{D2O} when compared to the DPPC/RaLPS model. This may initially seem surprising when referring to our previous hypothesis that the maximum information is obtainable in the contrast opposite to that of the tailgroup, e.g., \ce{D2O} for hydrogenated tailgroups. However, we need to consider that there will be important information in both \ce{D2O} and \ce{H2O} contrasts, but it will take more time to obtain the equivalent complementary information in the contrast matching the tailgroup deuteration state when compared to the contrast with maximum variation. We see that this is indeed the case when looking at the optimal time splits for two contrasts; the contrast with the most information (i.e., that which is chosen when only a single contrast can be measured) does not need to be measured as long as the optimal second choice. From this, we suggest that an optimal ``model blind'' measurement strategy would be to measure two contrasts (\ce{D2O} and \ce{H2O}), and either spend an equal time on both, or spend slightly longer on the contrast matching the tailgroup deuteration state.

\begin{table}
\centering
\begin{tabular}{@{\extracolsep{4pt}}cccccccccc}
\hline \hline
\multirow{2}{*}{Sample}      & \multicolumn{4}{c}{Contrast SLD ($10^{-6} \text{\AA}^{-2}$)} & \multicolumn{4}{c}{Split of Time (\%)} & \multirow{2}{*}{\begin{tabular}[c]{@{}c@{}}Minimum \\ Eigenvalue\end{tabular}} \\ \cline{2-5} \cline{6-9}
                             & 1st            & 2nd           & 3rd           & 4th          & 1st       & 2nd     & 3rd     & 4th   &                        \\ \hline
\multirow{4}{*}{DMPC}        & 6.36           & -             & -             & -            & 100.0     & -       & -       & -     & $4.957 \times 10^{-6}$ \\
                             & -0.56          & 6.36          & -             & -            & 70.7      & 29.3    & -       & -     & $1.919 \times 10^{-3}$ \\
                             & -0.56          & 1.91          & 6.36          & -            & 22.0      & 26.5    & 51.5    & -     & $2.666 \times 10^{-3}$ \\ 
                             & -0.56          & 1.90          & 1.93          & 6.36         & 22.7      & 5.8     & 19.4    & 52.1  & $2.666 \times 10^{-3}$ \\ \hline
\multirow{4}{*}{DPPC/RaLPS}  & -0.56          & -             & -             & -            & 100.0     & -       & -       & -     & $6.367 \times 10^{-8}$ \\
                             & -0.56          & 6.36          & -             & -            & 21.9      & 78.1    & -       & -     & $2.771 \times 10^{-3}$ \\
                             & -0.56          & 3.38          & 6.36          & -            & 24.5      & 21.5    & 54.0    & -     & $4.387 \times 10^{-3}$ \\
                             & -0.56          & 3.37          & 3.41          & 6.36         & 25.0      & 3.8     & 17.7    & 53.6  & $4.387 \times 10^{-3}$ \\
\hline \hline
\end{tabular}
\caption{Optimised contrast SLDs and counting time splits for the DMPC and DPPC/RaLPS bilayer models using 1 to 4 contrasts. Also shown is minimum eigenvalue for each set of conditions which was maximised using DE optimisation.}
\label{tab:bilayer_contrasts}
\end{table}

Now that we have an idea about how many contrasts to measure, we can optimise the choice of angle(s). Table \ref{tab:bilayer_angles} shows the optimised angles and counting time splits for each bilayer model, assuming a fixed time-budget, using 1 to 4 angles, and \ce{D2O} and \ce{H2O} contrasts. From table \ref{tab:bilayer_angles}, it appears that measuring two angles is an improvement on just one but with any more angles, the choices are once more repeated. The results suggest that, for the DMPC and DPPC/RaLPS models, a single low angle ($0.8^{\circ}$ and $0.7^{\circ}$, respectively) and a single high angle ($4.0^{\circ}$) is preferable, and this is generally in line with current practices, although for OFFSPEC, angles of more than $3.0^{\circ}$ are not commonly measured. For reference, the original experimentally-measured data for the DPPC/RaLPS bilayer was measured using two angles of $0.7^{\circ}$ and $2.3^{\circ}$ \cite{Clifton2016}. A low angle typically captures detail about the critical edge whereas a high angle provides information at high $Q$. What may come as a surprise is the counting times with only about 11\% and 9\% of the total time budget spent measuring the low angles in each case respectively. This could suggest that there is little information about the model parameters of interest at low $Q$ (there is no critical edge for \ce{H2O}) and it is therefore not worthwhile measuring for very long when compared to measuring for longer at a higher angle. Alternatively, it could indicate that the information at low $Q$ can be extracted more easily and therefore less time need be spent probing there. We note that if the times to change angle were included in the optimisation, the results could differ slightly from those shown here. This result is also instrument-dependent. For example, it is likely that a white beam instrument with a smaller wavelength range would favour more angles, although for any instruments similar to OFFSPEC (wavelength range of [1,14]\AA) the optimal angles will be very similar. We also note that it may be beneficial to run sub-optimal angles since increased overlap of data points may add diagnostic data, e.g. checking that the data scales correctly with angle and that overlapping data points agree.

\begin{table}
\centering
\begin{tabular}{@{\extracolsep{4pt}}ccccccccccc}
\hline \hline
\multirow{2}{*}{Sample}      & \multicolumn{4}{c}{Angle ($^{\circ}$)} & \multicolumn{4}{c}{Split of Time (\%)} & \multirow{2}{*}{\begin{tabular}[c]{@{}c@{}}Minimum\\ Eigenvalue\end{tabular}} \\ \cline{2-5} \cline{6-9}
                             & 1st      & 2nd      & 3rd     & 4th     & 1st       & 2nd      & 3rd     & 4th    &                        \\ \hline
\multirow{4}{*}{DMPC}        & 3.17     & -        & -       & -       & 100.0     & -        & -       & -      & $1.027 \times 10^{-2}$ \\
                             & 0.82     & 4.00     & -       & -       & 11.9      & 88.1     & -       & -      & $1.599 \times 10^{-2}$ \\
                             & 0.8     & 3.94     & 4.00    & -       & 11.8      & 2.9      & 85.3    & -      & $1.599 \times 10^{-2}$ \\ 
                             & 0.8     & 3.98     & 3.99    & 4.00    & 11.6      & 45.0      & 32.1    & 11.3    & $1.599 \times 10^{-2}$ \\ \hline
\multirow{4}{*}{DPPC/RaLPS}  & 2.91     & -        & -       & -       & 100.0     & -        & -       & -      & $1.035 \times 10^{-2}$ \\
                             & 0.68     & 4.00     & -       & -       & 8.9       & 91.1     & -       & -      & $2.314 \times 10^{-2}$ \\ 
                             & 0.68     & 4.00     & 4.00    & -       & 9.1       & 5.8      & 85.2    & -      & $2.313 \times 10^{-2}$ \\
                             & 0.43     & 0.69     & 4.00    & 4.00    & 0.0       & 8.7      & 61.4    & 29.9    & $2.313 \times 10^{-2}$ \\
\hline \hline
\end{tabular}
\caption{Optimised measurement angles and counting time splits for the DMPC and DPPC/RaLPS bilayer models using 1 to 4 angles. Also shown is minimum eigenvalue for each set of conditions which was maximised using DE optimisation.}
\label{tab:bilayer_angles}
\end{table}

\subsubsection{Underlayers}
As detailed in section \ref{bilayer_methods}, we were able to parameterise the bilayer models as a function of an added underlayer, whose thickness and SLD we can control. Table \ref{tab:bilayer_underlayers} shows the optimised underlayer properties for each bilayer model using 0 to 3 underlayers, assuming \ce{D2O} and \ce{H2O} contrasts are being measured. As can be seen, there is a large improvement in adding an underlayer to both models. There does also seem to be an improvement in adding multiple underlayers, but the relative improvements attained are comparatively small. When considering implementation of these designs in practice, two underlayers are typically required to achieve the desired surface chemistry \cite{Clifton2015}, but the additional underlayer parameters may result in the model becoming too complex for the experimentally-measured data. As a result, fitting could be problematic due to large parameter uncertainties and so, for simplicity, we proceed here with the addition of a single underlayer.

\begin{table}[h]
\centering
\begin{tabular}{@{\extracolsep{4pt}}cccccccc}
\hline \hline
\multirow{2}{*}{Sample}      & \multicolumn{3}{c}{Underlayer SLD ($10^{-6} \text{\AA}^{-2}$)} & \multicolumn{3}{c}{Underlayer Thickness ($\text{\AA}$)} & \multirow{2}{*}{\begin{tabular}[c]{@{}c@{}}Minimum\\ Eigenvalue\end{tabular}} \\ \cline{2-4} \cline{5-7}
                             & Layer 1             & Layer 2             & Layer 3             & Layer 1           & Layer 2           & Layer 3          &                        \\ \hline
\multirow{4}{*}{DMPC}        & -                   & -                   & -                   & -                 & -                 & -                & $4.694 \times 10^{-4}$ \\ 
                             & 5.39                & -                   & -                   & 127.1             & -                 & -                & $4.016 \times 10^{-3}$ \\ 
                             & 5.19                & 8.99                & -                   & 50.2              & 105.9             & -                & $5.284 \times 10^{-3}$ \\
                             & 2.32                & 5.34                & 8.99                & 29.3              & 46.8              & 129.4            & $5.414 \times 10^{-3}$ \\ \hline
\multirow{4}{*}{DPPC/RaLPS}  & -                   & -                   & -                   & -                 & -                 & -                & $1.312 \times 10^{-3}$ \\
                             & 9.00                & -                   & -                   & 76.5              & -                 & -                & $1.128 \times 10^{-2}$ \\
                             & 9.00                & 1.70                & -                   & 61.7              & 23.2              & -                & $1.818 \times 10^{-2}$ \\
                             & 1.01                & 8.93                & 2.33                & 61.0              & 63.1              & 15.8             & $1.944 \times 10^{-2}$ \\
\hline \hline
\end{tabular}
\caption{Optimised underlayer SLDs and thicknesses for the DMPC and DPPC/RaLPS bilayer models using 0 to 3 underlayers. Also shown is minimum eigenvalue for each set of underlayer properties which was maximised using DE optimisation. Layer 1 is nearest to the silicon substrate and layer 3 is nearest to the water solution.}
\label{tab:bilayer_underlayers}
\end{table}

\begin{figure}
\centering
\includegraphics[width=0.85\textwidth]{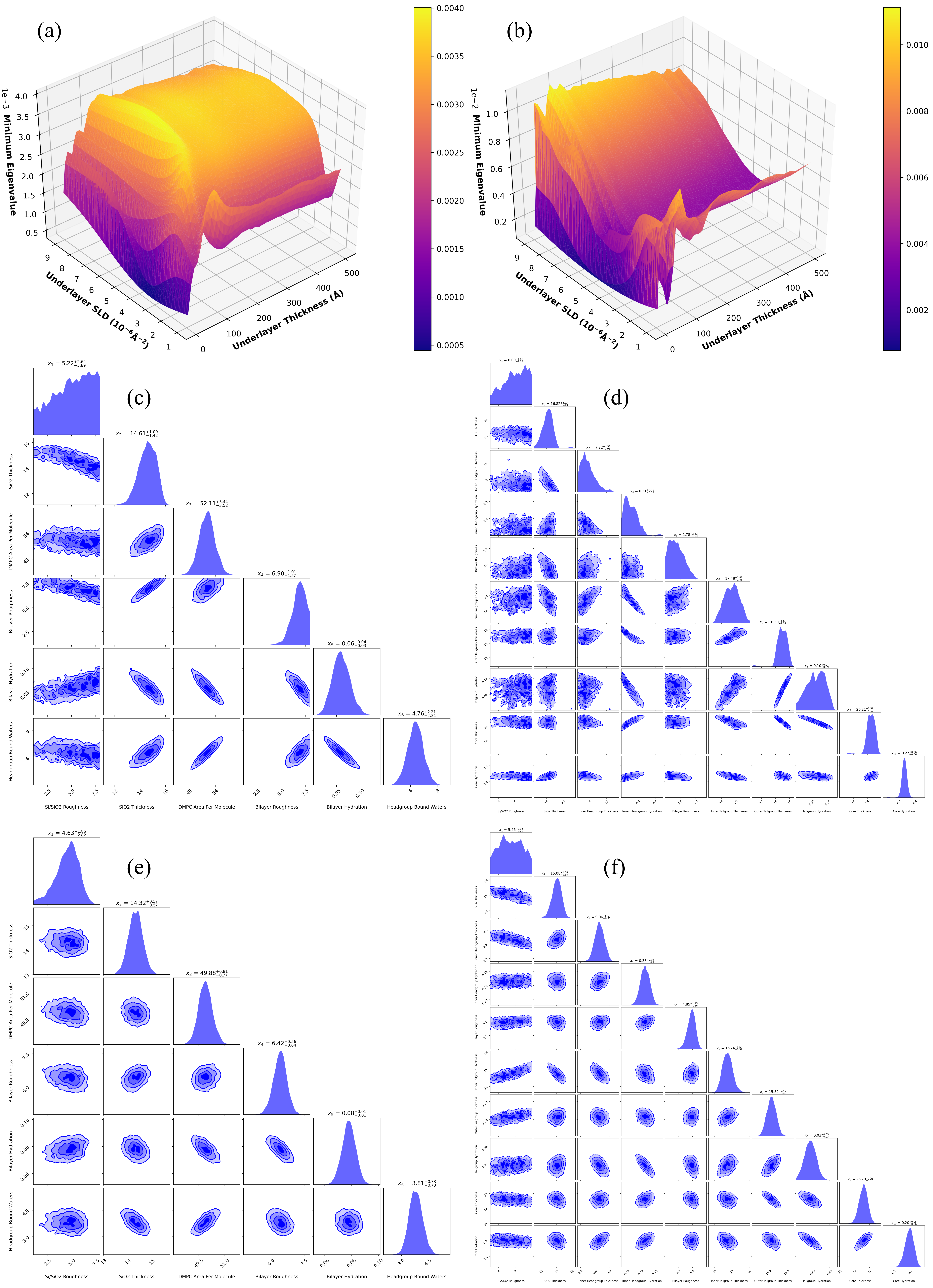}
\caption{Plots on the left correspond to the DMPC bilayer and plots on the right to the DPPC/RaLPS bilayer. Shown (\texttt{a} and \texttt{b}) are the plots of minimum eigenvalue versus underlayer SLD and underlayer thickness for the bilayer models, assuming \ce{D2O} and \ce{H2O} are being measured. Also shown are the nested sampling corner plots from sampling simulated data of \ce{D2O} and \ce{H2O} without any underlayer (\texttt{c} and \texttt{d}) and with a single underlayer (\texttt{e} and \texttt{f}) defined by the optimised values of table \ref{tab:bilayer_underlayers}.}
\label{fig:bilayer_underlayers}
\end{figure}

We compare the values of table \ref{tab:bilayer_underlayers} to those of commonly used gold and Permalloy, whose SLD's are $4.7 \times 10^{-6} \text{\AA}^{-2}$ and $8.4 \times 10^{-6} \text{\AA}^{-2}$ respectively, and we see that these are relatively close to the optimal values for the DMPC experiments (albeit in a different order). Since most of the improvement comes from the addition of a single layer, we hypothesise that the majority of measurements made on silicon would improve if measured with a gold or Permalloy underlayer, even though this is not the optimal underlayer for all circumstances. For the DMPC bilayer, we get minimum eigenvalues of $3.835 \times 10^{-3}$ and $3.490 \times 10^{-3}$ for the addition of single 100\AA\ gold and Permalloy underlayers respectively (compared to $4.694 \times 10^{-4}$ with no underlayer); for the DPPC/RaLPS bilayer, we get minimum eigenvalues of $4.576 \times 10^{-3}$ and $1.011 \times 10^{-2}$ respectively (compared to $1.312 \times 10^{-3}$ with no underlayer). Therefore, for DMPC, it appears that either gold or Permalloy will result in a significant improvement but for DPPC/RaLPS, Permalloy is preferable. To better visualise the optimisation space, figure \ref{fig:bilayer_underlayers} shows how the minimum eigenvalue changes for each SLD and thickness of the added underlayer. As can be seen, for the DMPC bilayer, the majority of optimisation space is relatively flat and so most underlayer SLDs and thicknesses will result in a significant improvement. However, for the DPPC/RaLPS bilayer, this is not the case, and the choice of both conditions appears to have a more significant effect.

To illustrate the improvement of the underlayer addition, nested sampling was used on simulated \ce{D2O} and \ce{H2O} data with the default \texttt{dynesty} stopping criteria, as shown in figure \ref{fig:bilayer_underlayers}. The simulation conditions were the same as previously used except only the $0.7^{\circ}$ angle was simulated to better illustrate the improvement in parameter uncertainties and posterior distributions with the added underlayer; with two angles, the distributions are already well-defined (as can be seen in figure \ref{fig:bilayer_contrasts}) and so the improvement would be less noticeable. From figure \ref{fig:bilayer_underlayers}, an improvement in both the estimated posterior distributions and 95\%/2-sigma credible intervals can clearly be seen when the suggested underlayer is added.

\subsection{Kinetics}
\begin{figure}
\centering
\includegraphics[width=0.55\textwidth]{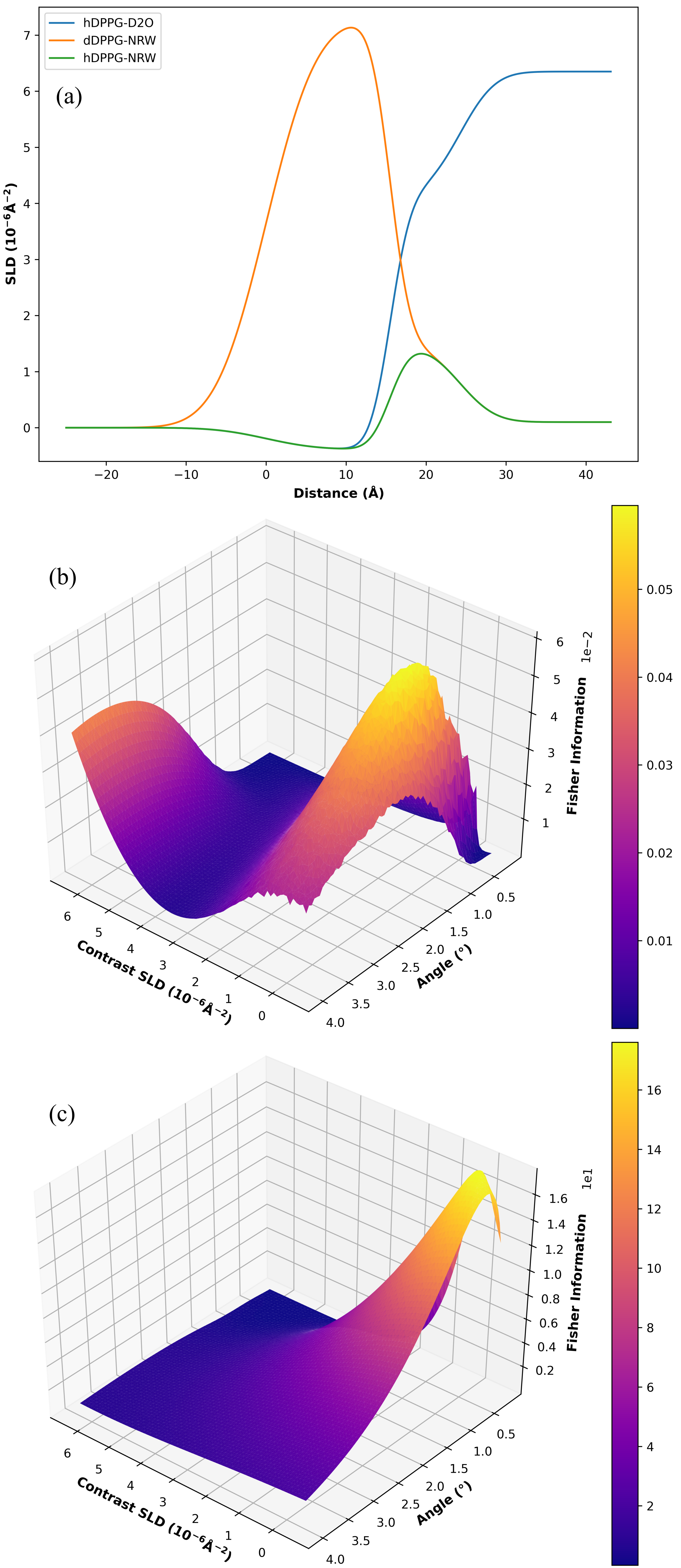}
\caption{Shown are the model SLD profiles for the d-DPPG monolayer on NRW, h-DPPG on NRW and h-DPPG on \ce{D2O} (\texttt{a}). Also shown are the plots of Fisher information in the lipid APM versus contrast SLD and measurement angle for the h-DPPG monolayer (\texttt{b}) and d-DPPG monolayer (\texttt{c}) models.}
\label{fig:kinetics}
\end{figure}

Using the monolayer model of section \ref{kinetics_methods}, the lipid APM was increased (and therefore surface excess decreased) to simulate the monolayer degrading over time. Figure \ref{fig:kinetics} shows SLD profile for the monolayer and how the FI in the lipid APM parameter changes with measurement angle and contrast SLD. The results were obtained for two models: one with h-DPPG and the other with d-DPPG.

For both h-DPPG and d-DPPG, a contrast SLD of approximately $0 \times 10^{-6} \text{\AA}^{-2}$ appears to maximises the APM information. This strongly agrees with common practices where NRW (SLD of approximately $0 \times 10^{-6} \text{\AA}^{-2}$) is commonly measured for its sensitivity to the water-lipid interfacial region \cite{Clifton2011}. When considering angles, $1.4^{\circ}$ and $0.4^{\circ}$ maximise the APM information in the h-DPPG and d-DPPG models respectively. For the d-DPPG model, there does also appear to be another solution that is only slightly worse than the optimal solution where \ce{D2O} is measured at a high angle. It is worth pointing out that there is more information in the lipid APM parameter when the tailgroups are deuterated, likely due to the increased reflectivity resulting in a greater total neutron count. 

\subsection{Magnetism}
\begin{figure}
\centering
\includegraphics[width=0.85\textwidth]{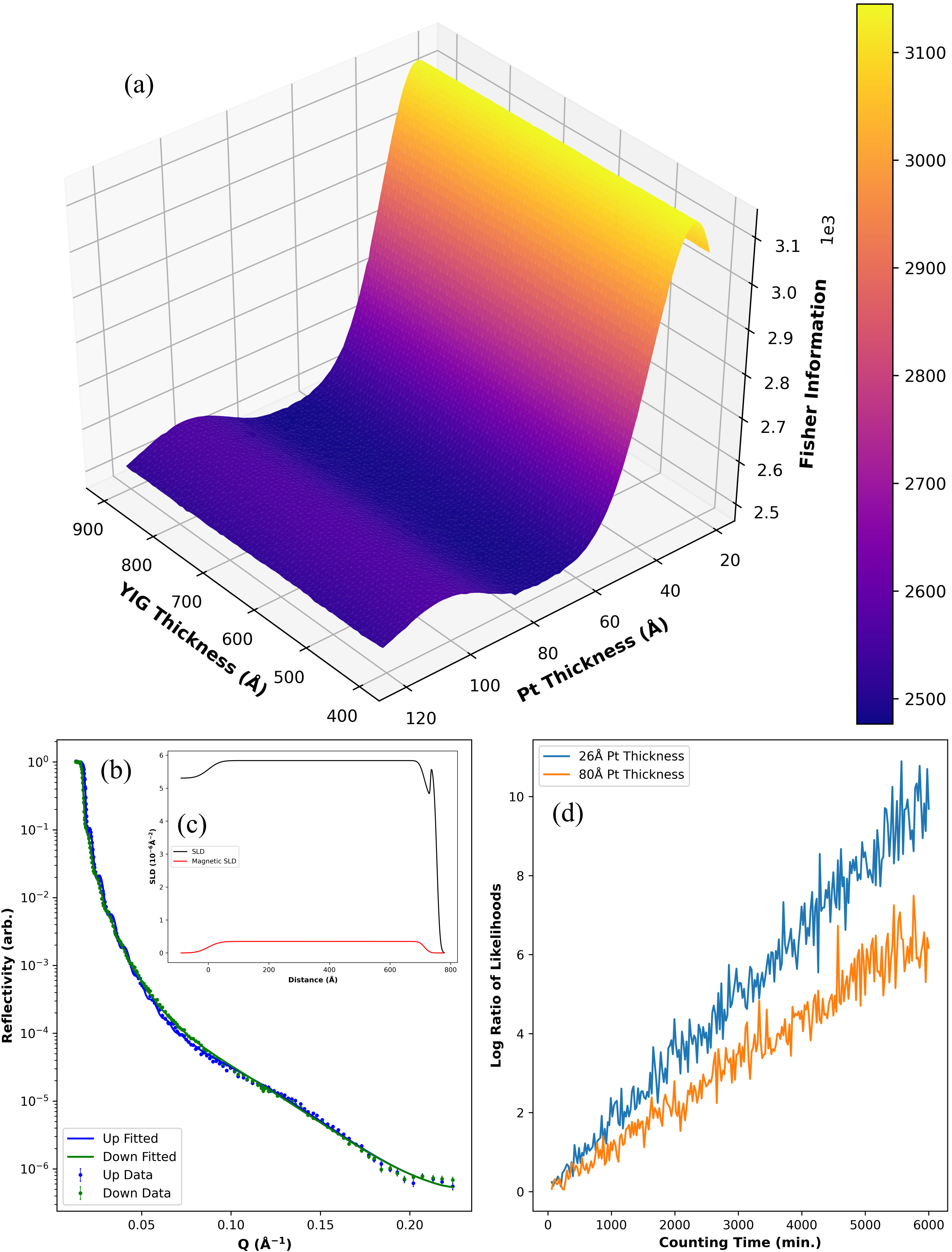}
\caption{Fisher information in the platinum layer magnetic SLD versus YIG and platinum layer thicknesses (\texttt{a}). Also shown are the experimentally-fitted SLD and reflectivity profiles (\texttt{b} and \texttt{c} respectively). Finally, shown is the log ratio of likelihoods between two models, one with an induced moment in the platinum layer and one with no moment, versus counting time (\texttt{d}) for two structures: one with an optimised 26\AA\ platinum layer thickness and the other with a sub-optimal 80\AA\ thickness. The counting times shown are minutes per measured spin state (i.e., for this model, the total times are double the values shown here).}
\label{fig:magnetism}
\end{figure}

Figure \ref{fig:magnetism} shows the experimentally-fitted SLD profile and reflectivity data for the magnetic sample of section \ref{magnetism_methods}. Also shown is how the FI in the platinum layer magnetic SLD changes with YIG and platinum layer thicknesses. It can be seen that the YIG thickness has virtually no effect on the FI in the platinum layer magnetic SLD. This is perhaps unsurprising, since the oscillations in reflectivity from the YIG are mostly damped at high $Q$ where the signal from the induced magnetism would lie. However, the platinum thickness does have an effect, and maximises the FI at around 26\AA. This is reasonably close to the fitted value of 21\AA\ (from the experimentally-measured data) and so, the original experiment appears to have fortuitously been near-optimal, given the varying conditions (i.e., ignoring the choice of angles, counting times etc.). It is worth noting that the difference in maximum and minimum FI in the figure \ref{fig:magnetism} plot is not particularly large and so, in practice, the improvement could be diminished by other unaccounted-for factors.

Figure \ref{fig:magnetism} demonstrates the improvement obtained using the suggested platinum layer thickness by illustrating how the log ratio of likelihoods between two models, one with an induced moment of 0.01 $\mu_{B}\slash \text{atom}$ and one with no moment, changes as a function of measurement time from 1 to 100 hours. The plot shows two lines: one for the optimised design with 26\AA\ platinum layer thickness and the other for a sub-optimal design with 80\AA\ platinum layer thickness. As can be clearly seen, over the entire range of times under consideration, the improved design requires a lower level of counting statistics to discriminate the 0.01 $\mu_{B}\slash \text{atom}$ moment. In fact, the change of thickness away from the optimal requires that the experiment be counted for several thousand more minutes to obtain the same certainty in the result. These results also show nicely that, with an optimised experiment, a moment of 0.01 $\mu_{B}\slash \text{atom}$ in a layer around only 20\AA\ thick is measurable in a reasonable time frame.

\subsection{Discussion}
This work has shown that, with relatively minor tweaks to experimental procedures, experiments can be performed in shorter time frames or with greater confidence in the results of interest. In covering the wide breath of experimental techniques presented here, we have made occasional assumptions that may result the optimal conditions presented here not necessarily being those that you might choose to measure with, e.g., changing contrast has no time penalty, or measuring higher angles do not increase background significantly. We have shown that the optimal values for a large number of the variable experimental measurement conditions are highly model-dependent. However, taking the example of the two bilayer models, while the optimal third contrast choice is quite different, a commonly chosen middle value of SMW avoids the complications of having to know the model in advance, with only a very minor loss of information. The shape of many of the optimisation surfaces are complex, and it may well suit the experimenter better to choose a slightly less optimal, but flatter region for experimental design, for example if the behaviour is highly oscillatory. For multi-parameter problems where larger dimensional spaces are to be probed differential evolution is able to provide the global optimum without exhaustive calculation, however, in these cases care should be taken to avoid sharp maxima if not all parameters can be appropriately controlled.

We note that the increases in minimum eigenvalue shown here correspond to better experimental outcomes; an increase by a factor of 100 in minimum eigenvalue corresponds to a reduction in uncertainties of a factor of 10 across the worst linear combination of parameters. Additionally, since the FI increases linearly with neutron counts, an increase in the information gained is equivalent to a commensurate increase in the neutron flux. As was shown with the magnetic example, this can mean the difference between an experiment being feasible in the allotted time, or not. We should note that the maximin optimisation we use here is not the only possible way of reducing the $N \times M$ matrix of the FI down to a single number (and changing this in the codebase is fairly trivial), but it does represent what we believe would be the "expected" optimal, i.e. minimising variance \textit{and} co-variance across all parameters of interest.

The FI framework is developed from a frequentist view, but similar Bayesian approaches have also been developed, also aimed at reflectometry experiment optimisation. As described in our previous work \cite{Durant2021}, the main difference in approaches is that the Fisher information is calculated from the model and the neutron counts, this therefore assumes the model describes the data, but gives the benefit of never having to fit the data or sample the posteriors. The Bayesian approach instead probes the data, trying to match models to the data and reconstruct the posteriors, requiring, at the very least, fitting of the data sets and expensive calculations which would usually preclude the development of equivalent tools to those presented here. One such Bayesian approach quantifies the extractable information gain by comparing the entropies of the prior and posterior probability density functions, representing the knowledge about a sample before and after an experiment \cite{Treece2019, Heinrich2020}. They draw many of the same conclusions which are presented here about contrasts and underlayers. However, the framework requires the use of computationally-expensive MCMC simulation and therefore may not be suitable for on-experiment design optimisation or searching large parameter spaces.

The code for interfacing the framework, with all of the model systems presented in this work, is available in our GitHub repository \cite{GitHub}. The repository also presents many of the examples discussed here as interactive Jupyter notebooks. These examples should be relatively easily modifiable for local applications. An incident flux file profile for the instrument being simulated is required, however the files for several of the ISIS instruments are provided.

\section{Conclusions}
In this work, we demonstrated how the Fisher information can be used to optimise experimental design across a wide range of different scientific applications in neutron reflectometry. We have shown that, for the two lipid bilayer systems investigated, the addition of an underlayer on the silicon substrate could provide a significant improvement, often over a factor of 10 in the minimum eigenvalue, but that the details of the optimal underlayer are model-dependent. When choosing contrasts in these experiments, the first two should always be pure \ce{D2O} and \ce{H2O}; the optimal third contrast is model-dependent but the commonly used silicon matched water is a good compromise if the model is not well-known. For a kinetic measurement of a monolayer on water, we showed that the optimal water contrast to measure is air-matched, but that the optimal angle depends on the deuteration state of the monolayer. For magnetic measurements, we demonstrated that it is possible to measure induced moments as low as 0.01 $\mu_{B}\slash \text{atom}$ in a thin layer, as long as the sample is optimised, and that it is possible to determine how long the measurement should be conducted in order to exceed any given confidence threshold that the moment is present.

The framework is not specific to neutron reflectometry; any technique that can be accurately simulated, and whose error bars rely on Poisson statistics, can interface with the code. A number of Jupyter notebooks as well as all of the code for this work is open-source and freely available on GitHub \cite{GitHub}.

\section{Acknowledgements}
We thank Luke Clifton for his assistance and expertise in fitting the lipid monolayer and lipid bilayer data sets.

{\small{\bibliography{main}}}

\section{Supporting Information}
\subsection{Experimental Design Optimisation}
\subsubsection{Importance Scaling}
We quantify parameter importance by choosing a suitable range for each parameter based on physical constraints. These ranges are then mapped to $[0,1]$. In this way, all parameters are put into the same, non-informational, importance units: a measure of information per importance. Such a linear mapping is analogous to a uniform prior in Bayesian statistics.

To transform the parameters of the Fisher information (FI) matrix, $\mathbf{g}^\xi$, we calculate $\mathbf{J^T}\mathbf{g}^\xi\mathbf{J}$, where $\mathbf{J}$ is the Jacobian for the transform. Since the importance mapping relationship is linear, the transform will be of the form $y_i = m_i x_i + c_i$ and $\mathbf{J}$ will be a diagonal matrix with entries $m_i$, i.e., each entry of the FI matrix, $\mathbf{g}^\xi_{i,j}$, is scaled by $m_i$ times $m_j$. For example, if parameter $i$ is 10 times more important than parameter $j$, then the $m_i$ will be 10 times $m_j$. Say the specified interval for a parameter, $x$, is $[a,b]$, then the linear transform to $y$, with interval $[0,1]$, is of the form 
$$y = (x - a)/(b - a) = x / (b - a) - a / (b - a)$$
In the form $y = m x + c$, we have $m = 1/(b - a)$ and $c = a / (b-a)$, and hence the required values for the coordinate transform.

\subsubsection{Maximin Optimisation}
\begin{figure}
\centering
\includegraphics[width=0.8\linewidth]{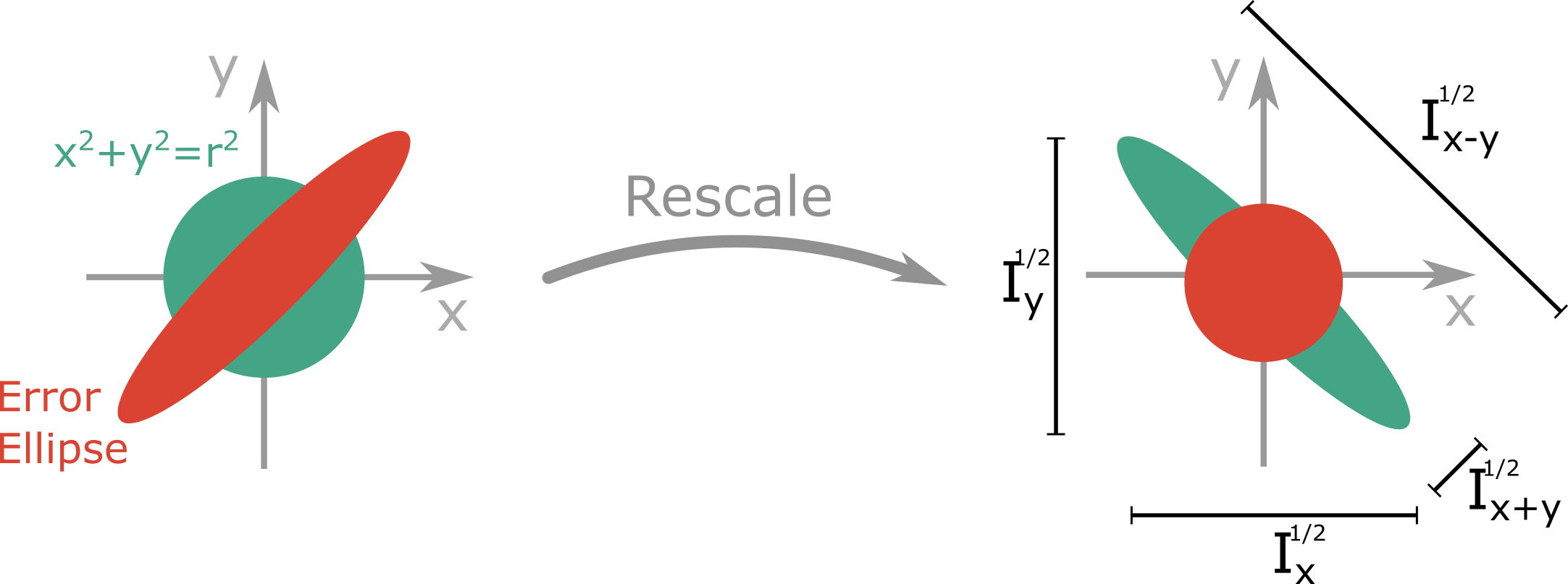}
\caption{Shown is an error ellipse with a large relative error in a combination of parameters (the $x+y$ combination). Although there is information in both $x$ and $y$, there is relatively little in the $x+y$ direction. 
\textit{Details:}
If we linearly transform the space on the left so that the error ellipse becomes the unit circle, the unit circle will be transformed into an inverted form of the error ellipse with sizes corresponding to the information content.}
\label{fig:ellipse}
\end{figure}

The presence of information content in individual parameters may hide low information content in a combination of parameters. Our approach chooses the (linear) combination of values that (locally) has the least information content and maximises it. Figure \ref{fig:ellipse} (right) shows an ellipse representing the FI, and how the ellipse corresponds to the amount of information in each parameter. The eigenvector of the FI matrix with minimum eigenvalue corresponds to the minor (short) axis of the ellipse and will be the linear combination of values with the least information.

The details of the mathematics describing the information quantities in figure \ref{fig:ellipse} are as follows: consider two pairs of nearby parameters, one represented by a point in 2D parameter space $p$ and another by $p + \Delta p$, where the magnitude of $\Delta p$ is small and equal to $r$. 
The information divergence between these points is given by
\[ D\left(p \,\Vert\, p + \Delta p\right) = \frac{1}{2} \sum_{ij} g_{ij} \Delta p_i \Delta p_j + O\left(\left\lVert \Delta p\right\rVert^3\right)\]
If $\Delta p$ is in the direction of the $x$-axis we can get an approximation which we will call $I_x$
\[ D\left(p \,\Vert\, p + \Delta p\right) \approx \frac{1}{2} r^2 g_{xx} = I_{x} \]
Similarly, in the $y$ direction we have $I_y = r^2 g_{yy}/2$, and in the $45^{\circ}$ directions, $x+y$ and $x-y$, we have 
\[I_{x+y} = \frac{1}{4}r^2(g_{xx} + g_{yy}) + \frac{1}{2}r^2g_{xy} = \frac{1}{2}\left(I_x + I_y + r^2g_{xy}\right)\]
\[I_{x-y} = \frac{1}{2}\left(I_x + I_y - r^2g_{xy}\right)\]
From this we can see that the total information is the same, whether we use a basis of $x$ and $y$, or of $x+y$ and $x-y$:
\[I_x + I_y = I_{x+y} + I_{x-y}\]
But we also see that, if the off-diagonal entry of $g$ (i.e., $g_{xy}$) is either very positive or very negative, the information about $x+y$ or $x-y$ might be very small. 

\subsection{Parameterisations and Fitting}
\begin{table}[H]
\centering
\begin{tabular}{llc}
\hline \hline
Sample                               & Model Parameter               & Fitted Value        \\ \hline
\multirow{8}{*}{DMPC Bilayer}        & \ce{Si}/\ce{SiO2} Roughness   & 2.0\AA              \\ 
                                     & \ce{SiO2} Thickness           & 14.7\AA             \\ 
                                     & \ce{SiO2}/DMPC Roughness      & 2.0\AA              \\ 
                                     & \ce{SiO2} Hydration           & 24.5\%              \\ 
                                     & DMPC Area Per Molecule        & $49.9 \text{\AA}^2$ \\  
                                     & Bilayer Roughness             & 6.6\AA              \\ 
                                     & Bilayer Hydration             & 7.4\%               \\  
                                     & Headgroup Bound Waters        & 3.59                \\ \hline
\multirow{13}{*}{DPPC/RaLPS Bilayer} & \ce{Si}/\ce{SiO2} Roughness   & 5.5\AA              \\ 
                                     & \ce{SiO2} Thickness           & 13.4\AA             \\ 
                                     & \ce{SiO2}/Bilayer Roughness   & 3.2\AA              \\ 
                                     & \ce{SiO2} Hydration           & 3.8\%               \\ 
                                     & Inner Headgroup Thickness     & 9.00\AA             \\  
                                     & Inner Headgroup Hydration     & 39.0\%              \\ 
                                     & Bilayer Roughness             & 4.0\AA              \\ 
                                     & Inner Tailgroup Thickness     & 16.7\AA             \\ 
                                     & Outer Tailgroup Thickness     & 14.9\AA             \\ 
                                     & Tailgroup Hydration           & 0.9\%               \\
                                     & Core Thickness                & 28.7\AA             \\ 
                                     & Core Hydration                & 26.0\%              \\ 
                                     & Asymmetry Value               & 0.95                \\ \hline
\multirow{5}{*}{DPPG Monolayer}      & Air/Tailgroup Roughness       & 5.0\AA              \\
                                     & Tailgroup/Headgroup Roughness & 2.0\AA              \\
                                     & Headgroup/Water Roughness     & 3.5\AA              \\
                                     & Lipid Area Per Molecule       & $54.1 \text{\AA}^2$ \\
                                     & Headgroup Bound Waters        & 6.69                \\
\hline \hline
\end{tabular}
\caption{Model parameters and associated fitted values for the DMPC bilayer, DPPC/RaLPS bilayer and DPPG monolayer models.}
\label{tab:lipids_fit}
\end{table}

\subsubsection{Lipid Bilayers}
Details of the model parameterisation for the 1,2-dimyristoyl-\textit{sn}-glycero-3-phosphocholine (DMPC) bilayer model can be found in our previous work and will not be repeated here. Table \ref{tab:lipids_fit} summarises the fitted parameters of the model. For the 1,2-dipalmitoyl-\textit{sn}-glycero-3-phosphocholine (DPPC)/\ce{Ra} lipopolysaccharide (LPS) bilayer model, the level of instrument background for the \ce{D2O} ($6.14 \times 10^{-6} \text{\AA}^{-2}$), silicon-matched water ($2.07 \times 10^{-6} \text{\AA}^{-2}$) and \ce{H2O} ($-0.56 \times 10^{-6} \text{\AA}^{-2}$) data were $4.6 \times 10^{-6}$, $8.6 \times 10^{-6}$ and $8.7 \times 10^{-6}$ respectively. Each experimental data set was fitted with an instrument resolution function of constant 4\% $dQ/Q$ and an experimental scale factor of 0.8. Table \ref{tab:lipids_fit} summarises the fitted parameters of the model. The DPPC/RaLPS bilayer model was defined using a slab representation: silicon, silicon oxide, inner headgroup, inner tailgroup, outer tailgroup, LPS core region and finally the bulk water solution of given scattering length density (SLD), $\rho_{\text{water}}$.

The model was defined with three roughness parameters: the silicon/silicon oxide and silicon oxide/bilayer interfacial roughnesses and a bilayer roughness that was shared between the other interfaces (inner headgroup/inner tailgroup, inner tailgroup/outer tailgroup, outer tailgroup/LPS core and LPS core/solution). The silicon substrate layer was defined using the known SLD of silicon ($2.07 \times 10^{-6} \text{\AA}^{-2}$). The silicon oxide and inner headgroup layers were defined using their known SLDs ($3.41 \times 10^{-6} \text{\AA}^{-2}$ and $1.98 \times 10^{-6} \text{\AA}^{-2}$ respectively), with the thickness and hydration of each layer set as parameters. The inner and outer tailgroup layers were defined using separate thickness parameters but a shared hydration parameter. The SLDs for the two layers, $\rho_{\text{inner}_{\text{TG}}}$ and $\rho_{\text{outer}_{\text{TG}}}$, were defined using an asymmetry parameter, $\alpha$, and the known SLDs of the DPPC and LPS tailgroups ($7.45 \times 10^{-6} \text{\AA}^{-2}$ and $-0.37 \times 10^{-6} \text{\AA}^{-2}$ respectively).
$$\rho_{\text{inner}_{\text{TG}}} = \alpha \rho_{\text{DPPC}_{\text{TG}}} + (1-\alpha) \rho_{\text{LPS}_{\text{TG}}}$$
$$\rho_{\text{outer}_{\text{TG}}} = (1-\alpha) \rho_{\text{DPPC}_{\text{TG}}} + \alpha \rho_{\text{LPS}_{\text{TG}}}$$

The LPS core layer thickness and hydration were set as parameters, but the SLD was defined using the mole fraction of \ce{D2O} from the bulk water SLD, denoted here as $x$, and the known SLDs of the LPS core in \ce{D2O}, $\rho_{\text{core}_{\text{\ce{D2O}}}}$, and \ce{H2O}, $\rho_{\text{core}_{\text{\ce{H2O}}}}$ ($4.20 \times 10^{-6} \text{\AA}^{-2}$ and $2.01 \times 10^{-6} \text{\AA}^{-2}$ respectively).
$$x = \frac{\rho_{\text{water}}-\rho_{\text{\ce{H2O}}}}{\rho_{\text{\ce{D2O}}}-\rho_{\text{\ce{H2O}}}}$$
$$\rho_{\text{core}} = x\rho_{\text{core}_{\text{\ce{D2O}}}} + (1-x)\rho_{\text{core}_{\text{\ce{H2O}}}}$$

\vspace{-1em}
\subsubsection{Kinetics}
The 1,2-dipalmitoyl-sn-\textit{glycero}-3-phospho-(1-rac-glycerol) (DPPG) monolayer model was defined using a slab representation: air, monolayer tailgroup (either hydrogenated or deuterated), monolayer headgroup, and finally the bulk water solution of given SLD, $\rho_{\text{water}}$; table \ref{tab:lipids_fit} summarises the parameters of the model. All model interfacial roughnesses (air/tailgroup, tailgroup/headgroup and headgroup/water) were parameterised. The tailgroup (both hydrogenated and deuterated) and headgroup thicknesses were defined using a shared lipid area per molecule (APM) parameter and the equation ${d} = \text{V}/\text{A}$, where $d$ is the layer thickness, $V$ is the layer volume and A is the lipid APM.

The tailgroup and headgroup volumes were calculated from the volumes of their constituent components, as summarised in table \ref{tab:DPPG_vols}. For the tailgroup volume, $V_{\text{TG}}$, this was relatively straightforward
$$V_{\text{TG}} = 28V_{\text{\ce{CH2}}} + 2V_{\text{\ce{CH3}}}$$
but for the headgroups, we needed to account for the hydrating water molecules. We did this by first multiplying the known water volume by the headgroup bound waters parameter to obtain the extra water volume in the headgroups, $V_{\text{bound}}$, and then added this to the individual fragment volumes
$$V_{\text{HG}} = V_{\text{\ce{PO4}}} + 2V_{\text{\ce{C3H5}}} + 2V_{\text{\ce{CO2}}} + V_{\text{bound}}$$

We calculated the hydrogenated and deuterated monolayer tailgroup SLDs using the previously calculated tailgroup volume and equation $\rho = \Sigma b / \text{V}$, where $\rho$ is the layer SLD, $\Sigma b$ is the neutron scattering length (SL) sum for the layer and $V$ is the layer volume. The SL sums of the hydrogenated and deuterated tailgroups, $\Sigma b_{\text{hTG}}$ and $\Sigma b_{\text{dTG}}$ respectively, were calculated from the total SL of each constituent fragment. The SLs of the individual elements of the fragments are given in table \ref{tab:DPPG_SLs}.
$$\Sigma b_{\text{hTG}} = 28\Sigma b_{\text{\ce{CH2}}} + 2\Sigma b_{\text{\ce{CH3}}}$$
$$\Sigma b_{\text{dTG}} = 28\Sigma b_{\text{\ce{CD3}}} + 2\Sigma b_{\text{\ce{CD3}}}$$
        
Like the tailgroups, the headgroup SLD was determined using the previously calculated headgroup volume, the SL sums of the constituent fragments, and equation $\rho = \Sigma b / \text{V}$. However, as with the headgroup volume calculation, we needed to account for the hydrating water molecules. We did this by first calculating the mole fraction of \ce{D2O} from the bulk water SLD, $x$, to get the average SL sum per water molecule.
$$x = \frac{\rho_{\text{water}}-\rho_{\text{\ce{H2O}}}}{\rho_{\text{\ce{D2O}}}-\rho_{\text{\ce{H2O}}}}$$
$$\Sigma b_{\text{water}} = x\Sigma b_{\text{\ce{D2O}}} + (1-x)\Sigma b_{\text{\ce{H2O}}}$$

By multiplying this value by the headgroup bound waters parameter, we were able to obtain the SL sum of the hydrating water molecules, $\Sigma b_{\text{bound}}$, from which we could calculate the headgroup SL sum.
$$\Sigma b_{\text{HG}} = \Sigma b_{\text{\ce{PO4}}} + 2\Sigma b_{\text{\ce{C3H5}}} + 2\Sigma b_{\text{\ce{CO2}}} + \Sigma b_{\text{bound}}$$

\begin{table}[H]
    \begin{minipage}{0.48\textwidth}
        \centering
        \begin{tabular}{lc}
        \hline \hline
        Fragment  & Volume ($\text{\AA}^3$) \\ \hline
        \ce{CH2}  & 28.1                    \\
        \ce{CH3}  & 26.4                    \\
        \ce{CO2}  & 39.0                    \\
        \ce{C3H5} & 68.8                    \\
        \ce{PO4}  & 53.7                    \\
        Water     & 30.4                    \\
        \hline \hline
        \end{tabular}
        \caption{Volumes of the DPPG monolayer tailgroup and headgroup fragments.}
        \label{tab:DPPG_vols}
    \end{minipage}%
    \hfill
    \begin{minipage}{0.48\textwidth}
        \centering
        \begin{tabular}{lc}
        \hline \hline
        Component  & Scattering Length ($10^{-4} \text{\AA}$) \\ \hline
        Carbon     &  0.6646                                  \\
        Oxygen     &  0.5843                                  \\
        Hydrogen   & -0.3739                                  \\
        Phosphorus &  0.5130                                  \\
        Deuterium  &  0.6671                                  \\
        \hline \hline
        \end{tabular}
        \caption{Neutron scattering lengths for the components of the tailgroup and headgroup fragments.}
        \label{tab:DPPG_SLs}
    \end{minipage}
\end{table}

\vspace{-1em}
\subsubsection{Magnetism}
The experimental scale factor, level of instrument background and instrument resolution function used to fit the data were 1.025, $4 \times 10^{-7}$ and constant 2.8\% $dQ/Q$ respectively. The model was defined using a slab representation consisting of air, platinum, yttrium oxide, yttrium iron garnet (YIG) and yttrium aluminium garnet (YAG); table \ref{tab:YIG_fitted} summarises the fitted parameters.
\begin{table}[H]
\centering
\begin{tabular}{lcccc}
\hline \hline
Layer         & SLD ($10^{-6} \text{\AA}^{-2}$) & Thickness ($\text{\AA}$) & Roughness ($\text{\AA}$) & Magnetic SLD ($10^{-6} \text{\AA}^{-2}$) \\ \hline
Air           & 0.00                            & -                        & -                        & -                                        \\
Platinum      & 5.65                            & 21.1                     & 8.2                      & 0.00                                     \\
Yttrium Oxide & 4.68                            & 19.7                     & 2.0                      & -                                        \\
YIG           & 5.84                            & 713.8                    & 13.6                     & 0.35                                     \\
YAG           & 5.30                            & -                        & 30.0                     & -                                        \\
\hline \hline
\end{tabular}
\caption{Fitted SLD, thickness, roughness and magnetic SLD for each layer of the magnetic sample.}
\label{tab:YIG_fitted}
\end{table}

\end{document}